\documentstyle[epsfig,aps]{revtex}

\newcommand{\beq}{\begin{equation}}
\newcommand{\eeq}{\end{equation}}
\def\beqa{\begin{eqnarray}}
\def\eeqa{\end{eqnarray}}

\def\lap{\lower.5ex\hbox{$\; \buildrel < \over \sim \;$}}
\def\gap{\lower.5ex\hbox{$\; \buildrel > \over \sim \;$}}

\begin{document}
\draft

\twocolumn[\hsize\textwidth\columnwidth\hsize\csname
@twocolumnfalse\endcsname
\preprint{IHES/P/00/32\\  April 2000}
\date{  26 April 2000}

\title{Gravitational wave bursts from cosmic strings}

\author{Thibault Damour$^{1)}$ and Alexander Vilenkin$^{2)}$}

\address{$^{1)}$Institut des Hautes Etudes Scientifiques, F-91440 Bures-
sur-Yvette, France \\
$^{2)}$Physics Department, Tufts University, Medford, MA 02155, USA}

\maketitle
\begin{abstract}
Cusps of cosmic strings emit strong beams of high-frequency gravitational
waves (GW). As a consequence of these beams, the stochastic ensemble of
gravitational waves
generated by a cosmological network of oscillating loops is strongly
non Gaussian, and includes occasional sharp bursts that stand above
the rms GW background. These bursts might be detectable by the planned
GW detectors LIGO/VIRGO and LISA for string tensions as small as 
$G \mu \sim 10^{-13}$. The GW bursts discussed here might be accompanied
by Gamma Ray Bursts.
\end{abstract}
\vskip2pc]

Cosmic strings are linear topological defects that could be formed at
a symmetry breaking phase transition in the early universe.  Strings
are predicted in a wide class of elementary particle models and can
give rise to a variety of astrophysical phenomena \cite{Book}.  In
particular, oscillating loops of string can generate a potentially
observable gravitational wave (GW) background ranging over many
decades in frequency.  The spectrum of this stochastic background has
been extensively discussed in the literature \cite{V81,HR,VV,BB,CA,CBS},
but until now it has been tacitly assumed that the GW background is
nearly Gaussian.  In this paper, we show that the GW background from
strings is strongly non-Gaussian and includes sharp GW bursts (GWB)
emanating from cosmic string cusps \cite{BHV1}.  
We shall estimate the amplitude, frequency spectrum and rate of the bursts,
 and discuss their detectability by the planned GW detectors LIGO/VIRGO
and LISA.

We begin with a brief summary of the relevant string properties and 
evolution\cite{Book}.  A horizon-size volume at any cosmic time $t$
contains a few long strings stretching across the volume and a large
number of small closed loops. The typical length and number density of loops
 formed at time $t$  are approximately given by
\beq
l \sim \alpha t, ~~~~  n_l(t)\sim \alpha^{-1}t^{-3}.
\label{n}
\eeq
The exact value of the parameter $\alpha$ in (\ref{n}) is not known. We
shall assume, following \cite{BB}, that $\alpha$ is determined by the
gravitational backreaction, so that
$\alpha\sim \Gamma G\mu$,
where  $\Gamma\sim 50$ is a numerical coefficient, $G$ is
Newton's constant, and $\mu$ is the string tension, i.e. the
 mass per unit length of the string.
The coefficient $\Gamma$ enters the total rate of energy loss
by gravitational radiation
$ d{\cal E}/dt\sim \Gamma G\mu^2.$
For a loop of invariant length $l$
\cite{invlength}, the
oscillation period is $T_l=l/2$ and the lifetime is
$\tau_l\sim l/\Gamma G\mu \sim t $.

A substantial part of the radiated energy is emitted from
near-cusp regions where, for a short period of time, the string
reaches a speed very close to the speed of light \cite{VV}. 
 Cusps tend to be formed a few times
during each oscillation period \cite{T}. Let us estimate the (trace-reversed)
metric perturbation
$ \overline{h}_{\mu \nu} = h_{\mu \nu}- \frac{1}{2}h \eta_{\mu \nu}$
 emitted near a cusp. Let
  $ {\kappa}_{\mu \nu} = r_{\rm phys}\overline{h}_{\mu \nu}$ 
denote the product of the metric perturbation
 by the  distance away from the string,
estimated in the local wave-zone of the 
oscillating string. $ {\kappa}_{\mu \nu}$ is given by
a Fourier series whose coefficients are proportional to the Fourier
transform of the stress-energy tensor of the string:
 
\beq
T^{\mu \nu}(k^{\lambda}) = T_l^{-1} \int_{T_l} d\tau d\sigma \mu
( \dot{X}^{\mu} \dot{X}^{\nu} - X^{\prime \mu} X^{\prime \nu})
\exp(-i k.X).
\label{T(k)}
\eeq
Here, $X^{\mu}(\tau,\sigma)$ represents the string worldsheet, parametrized by
the conformal coordinates $\tau$ and $\sigma$. [$\dot{X} = {\partial}_{\tau}X$,
$X^{\prime} = {\partial}_{\sigma}X$.]
In the direction of emission $\bf n$, $k^{\mu} = (\omega,{\bf k})$ runs over the discrete
set of values $4 \pi l^{-1} m (1 , \bf n)$, where $m = 1,2,\ldots$. Near
a cusp (and only near a cusp) the Fourier series giving $ {\kappa}_{\mu \nu}$
is dominated by large $m$ values, and can be approximated by a 
continuous Fourier integral. The continuous Fourier component
(corresponding to an octave of frequency around $f$) 
 $ {\kappa}(f) \equiv |f| \widetilde{\kappa}(f) \equiv |f| 
 \int dt \exp( 2 \pi i f t) {\kappa}(t)$
 is then given by

\beq
{\kappa}_{\mu \nu}(f) = 2 G l |f| T_{\mu \nu}(k^{\lambda}).
\label{k(f)}
\eeq
In a conformal gauge, the string motion is given
by $X^{\mu}= \frac{1}{2}(X_{+}^{\mu}( {\sigma}_{+}) 
+ X_{-}^{\mu}({\sigma}_{-}))$,where ${\sigma}_{\pm} = \tau \pm \sigma$,
and where $\dot{X}^{\mu}_{\pm}$ is a null 4-vector. Further choosing
a``time gauge'' ($\tau = X^0 = t$ ) ensures that the time components 
of these null vectors are equal to 1. A cusp is a point on the worldsheet
where these two null vectors coincide, say 
$\dot{X}^{\mu}_{c \pm} = l^{\mu} = (1,\bf{n}_c)$. One can estimate the
waveform (\ref{k(f)}) emitted  near the cusp ( localized, say, at
${\sigma}_{+} ={\sigma}_{-} = 0$)  by replacing
 in the integral (\ref{T(k)}) $X_{\pm}^{\mu}$ by their local Taylor 
 expansions

\beq
X_{\pm}^{\mu}({\sigma}_{\pm}) = X_{c \pm}^{\mu} + l^{\mu}{\sigma}_{\pm}
         + \frac{1}{2}\ddot{X}^{\mu}_{\pm}{\sigma}_{\pm}^2
	 + \frac{1}{6}{\buildrel{\ldots}\over{X}}^{\mu}_{\pm}{\sigma}_{\pm}^3
	  + \ldots
\label{x}
\eeq 

For a given frequency $f \gg T_l^{-1}$,
 the integral (\ref{T(k)}) is significant only
if the angle $\theta$ between the direction of emission $\bf{n}$
and the ``3-velocity'' of the cusp $\bf{n}_c)$ is smaller than about
$ {\theta}_m \equiv (T_l|f|)^{-1/3}$. When $\theta \ll {\theta}_m$, the 
integral can be explicitly evaluated. After a suitable gauge transformation
one finds \cite{DV}

\beq
{\kappa}^{\mu \nu}(f) = - C G \mu (2\pi |f|)^{-1/3}
             A_{+}^{(\mu}A_{-}^{\nu )},
\label{waveform1}
\eeq
where $C= 4 \pi (12)^{4/3} ( 3 \Gamma(1/3))^{-2}$ and where the linear
polarization tensor is the symmetric tensor product of 
$A_{\pm}^{\mu} \equiv \ddot{X}^{\mu}_{\pm}/ |\ddot{X}_{\pm}|^{4/3}$.
The inverse Fourier transform of Eq.(\ref{waveform1}) gives a time-domain
waveform proportional to $ |t - t_c|^{1/3}$ , where $t_c$ corresponds to the
peak of the burst \cite{Vachaspati}. The sharp spike at $t=t_c$ exists 
only if $\theta =0$ (i.e. if one observes it exactly in the direction
defined by the cusp velocity). When $ 0 \neq \theta \ll 1$ the spike is
smoothed over $|t - t_c| \sim {\theta}^3 T_l $. In the Fourier domain
this smoothing corresponds to an exponential decay for frequencies
$ |f| \gg 1/(\theta^3T_l)$. 

Eq.(\ref{waveform1}) gives the waveform in the local wave-zone of the
oscillating loop:  $ \overline{h}_{\mu \nu} =
 {\kappa}_{\mu \nu} /r_{\rm phys} $. To take into account the subsequent
  propagation of this wave 
 over cosmological distances, until it reaches us,
 one must introduce three modifications in this waveform:	
 (i)  replace $r_{\rm phys}$ by $ a_0 r$ where $r$ is the comoving
 radial coordinate in a Friedman universe (taken to be flat:
 $ ds^2 = -dt^2 + a(t)^2 (dr^2 + r^2 d\Omega^2)$)
  and $a_0 = a(t_0)$ the present scale factor , (ii) express the locally
  emitted frequency in terms of the observed one 
  $f_{\rm em} = (1+z) f_{\rm obs}$
  where $z$ is the redshift of the source, and (iii) transport the
  polarization  tensor of the wave by parallel propagation (pp) along 
  the null geodesic   followed by the GW:

\beq
\overline{h}_{\mu \nu}(f) ={\kappa}_{\mu \nu}^{\rm pp}((1+z)f) /( a_0 r ).
\label{waveform2}
\eeq
Here, and henceforth, $f > 0$ denotes the observed frequency. In terms of
the redshift we have $ a_0 r = 3 t_0 ( 1 - (1+z)^{-1/2})$, where $t_0$
 is the present age of the universe (this relation holds during the matter
 era, and can be used for the present purpose in the radiation era 
 because $a_0 r$ has a finite limit for large $z$). 
 
 For our order-of-magnitude estimates we shall assume that 
 $ |\ddot{X}_{\pm}| \sim 2 \pi/l$. The various numerical factors 
 in the equations above nearly compensate each other to give 
 the following simple estimate for
 the observed waveform in the frequency domain
($ h(f) \equiv |f| \widetilde{h}(f) \equiv |f| \int dt \exp(2 \pi i f t) h(t)$)
 
 \beq
 h(f) \sim  \frac{G \mu l}{((1+z)fl)^{1/3}} \frac{1+z}{t_0 z}.
 \label{h(f)}
 \eeq
Here the  explicit redshift dependence is a convenient simplification of the
exact one given above.
This result holds only if, for a given observed frequency $f$,
the angle $\theta = \cos^{-1}({\bf n}\cdot{\bf n}_c)$ satisfies

\beq
 \theta \alt {\theta}_m \equiv ((1+z)fl/2)^{-1/3}.
 \label{theta}
\eeq 
To know the full dependence of $h(f)$ on the redshift we need to express
$ l \sim \alpha t$ in terms of $z$. We write

\beq
l \sim\alpha t_0 {\varphi}_{l}(z) ; ~~
{\varphi}_{l}(z) = (1+z)^{-3/2}(1+z/z_{eq})^{-1/2}.
 \label{l(z)}
\eeq
Here $z_{eq} \simeq 2.4 \times 10^4 \Omega_0 h_0^2 \simeq 10^{3.9}$
is the redshift of equal matter and radiation densities, and we found it
convenient to define the function $ {\varphi}_{l}(z)$ which interpolates
between the different functional  $z$-dependences of $l$ in the matter era,
and the radiation era. [We shall systematically introduce such interpolating
 functions of $z$, valid for all redshifts, in the following.]
  Inserting Eq.(\ref{l(z)}) into Eq.(\ref{h(f)}) yields
\beqa
h(f,z) \sim G \mu {\alpha}^{2/3} (ft_0)^{-1/3} {\varphi}_{h}(z),\nonumber \\ 
{\varphi}_{h}(z) = z^{-1} (1+z)^{-1/3}(1+z/z_{eq})^{-1/3}.
 \label{h(z)}
\eeqa

We can estimate the rate of GWBs originating at cusps in the redshift interval 
$dz$, and observed around the frequency $f$, as
$ d{\dot N} \sim  \frac{1}{4}{\theta}_m^2
(1+z)^{-1}\nu(z) dV(z)$.
Here, the first factor is the beaming fraction within the cone of 
maximal angle ${\theta}_m(f,z)$, Eq.(\ref{theta}), the second factor comes
from the relation $dt_{\rm obs}=(1+z)dt$,
$\nu(t)\sim f_c n_l(t)/T_l\sim 2\alpha^{-2}t^{-4}$ is the number of
cusp events per unit spacetime volume, $f_c\sim 1$ is the average number of
cusps per oscillation period of a loop, $T_l\sim \alpha
t/2$, and $dV(z)$ is the proper volume between
redshifts $z$ and $z+dz$. In the matter era
$dV=54\pi t_0^3[(1+z)^{1/2}-1]^2(1+z)^{-11/2} dz $, while in the radiation era 
$dV=72\pi t_0^3(1+z_{eq})^{1/2}(1+z)^{-5} dz $. The function
${\dot N}(f,z) \equiv  d{\dot N}/d \ln z$ can be approximately represented
by the following interpolating function of $z$ 

\beqa
{\dot N}(f,z)\sim10^2 t_0^{-1}{\alpha}^{-8/3} (ft_0)^{-2/3} {\varphi}_n(z),
\nonumber \\
{\varphi}_n(z) =  z^3 (1+z)^{-7/6} (1+z/z_{eq})^{11/6}.
\label{ndot}
\eeqa

The observationally most relevant question is:  what is the typical
 amplitude of bursts $h^{\rm burst}_{\dot N}(f)$that we can expect 
 to detect at some given rate
 $ {\dot N}$, say, one per year? Using ${\dot N} =\int_{0}^{z_m}{\dot N}(f,z)
 d \ln z \sim {\dot N}(f,z_m) $, where $z_m$ is the largest redshift 
 contributing to $ {\dot N}$, one can estimate 
$h^{\rm burst}_{\dot N}(f)$  by solving for $z$ in Eq.(\ref{ndot})
and substituting the result $z = z_m({\dot N},f)$
in Eq.(\ref{h(z)}). The final answer has a different functional form 
depending on the magnitude of the quantity
\beq
 y({\dot N} , f)\equiv 10^{-2} {\dot N} t_0 {\alpha}^{8/3} ( ft_0)^{2/3}. 
\eeq
Indeed, if 
$ y<1$ the dominant redshift will be $z_m(y) <1$; while, if 
$ 1<y< {z_{eq}}^{11/6}$, $ 1< z_m(y)< z_{eq}$, and if 
$ y > {z_{eq}}^{11/6}$, $ z_m(y) > z_{eq}$. We can again introduce a suitable
interpolating function $ g(y)$ to represent the final result as an
explicit function of $y$:

\beqa
h^{\rm burst}_{\dot N}(f) \sim 
G \mu {\alpha}^{2/3} (ft_0)^{-1/3} g[y({\dot N} , f)], \nonumber \\ 
g(y) = y^{-1/3} (1 + y)^{-13/33} ( 1+y/(z_{eq})^{11/6} )^{3/11}.
\label{hburst}
\eeqa

The prediction Eq.(\ref{hburst}) for the amplitude of the GWBs generated
at cusps of cosmic strings is the main new result of this work. 
To see whether or not these bursts can be
distinguished from the stochastic gravitational wave background, we
have to compare the burst amplitude (\ref{hburst}) to the rms amplitude of
the background, $h_{\rm rms}$, at the same frequency. We define $h_{\rm rms}$ 
 as the ``confusion'' part of the ensemble of
 bursts Eq.(\ref{hburst}), i.e. the superposition of all the ``overlapping''
 bursts, those whose occurrence rate is higher than their typical frequency.
 This is estimated by multiplying 
the square of Eq.(\ref{h(z)}) by the number of overlapping bursts within
a frequency octave 
$ n_z(f) \equiv f^{-1} {\dot N}(f,z)$ , and then integrating over all
$ \ln z$ such that $ n_z(f) > 1$ , and ${\theta}_m(f,z)< 1$:

\beq
h_{\rm rms}^{2}(f) = \int h^2(f,z) n_z(f) d \ln z  H(n_z -1)
H( 1 - {\theta}_m),
\label{hrms}
\eeq
where $H$ denotes Heaviside's step function.
Eq.(\ref{hrms}) differs from previous estimates of the stochastic background
\cite{V81,HR,VV,BB,CA,CBS} (beyond the fact that we use the simplified loop 
density model Eq.(\ref{n})) in that the latter did not incorporate the
 restriction to $ n_z(f) > 1$, i.e. they included non-overlapping bursts
 in the average of the squared GW amplitude.

It is easily checked from Eq.(\ref{hburst}) that $h^{\rm burst}$ is a 
 monotonically decreasing function of both ${\dot N}$ and $f$. These 
 decays can be described by (approximate) power laws , with an 
 index which depends on the relevant range of dominant redshifts; e.g.,
 as ${\dot N}$ increases,  $h^{\rm burst}$ decreases first like
 ${\dot N}^{-1/3}$ ( in the range $z_m<1$), then like ${\dot N}^{-8/11}$
 (when $1<z_m<z_{eq}$), and finally like
 ${\dot N}^{-5/11}$ (when $z_m>z_{eq}$). 
 For the frequency dependence of $h^{\rm burst}$ , the corresponding
 power-law indices are successively: $-5/9 , -9/11$ and $-7/11$.
 [These slopes come from combining the basic $f^{-1/3}$ dependence of
 the spectrum of each burst with the indirect dependence on $f$ of
 the dominant redshift $z_m(\alpha, \dot N , f)$.]
 By contrast, when using our assumed link
 $ G \mu \sim \alpha/50$ between the string tension $\mu$ and the parameter
 $\alpha$, one finds that the index of the power-law dependence of
 $h^{\rm burst}$ upon $\alpha$ takes successively the values
 $ +7/9, -3/11 $ and $ +5/11$. Therefore, in a certain range of values
 of $\alpha$ (corresponding to $1<z_m(\alpha,{\dot N},f)<z_{eq}$) 
 the GWB amplitude  (paradoxically) {\it increases} as one decreases
 $\alpha$ , i.e. $G\mu$.

In Fig. 1 we plot (as a solid line) the logarithm of the GW burst amplitude, 
$ {\log}_{10}(h^{\rm burst})$, as a function of 
$ {\log}_{10}(\alpha)$, for 
  ${\dot N} = 1~{\rm yr}^{-1}$, and for $f = f_c = 150$ Hz.
   This central frequency
  is the optimal one for the detection of a  $ f^{-1/3}$-spectrum burst
  by LIGO. We indicate on the same plot (as horizontal dashed lines)
  the (one sigma) noise levels
  $h^{\rm noise}$ of LIGO 1 (the initial detector), and LIGO 2 (its planned
  advanced configuration). The VIRGO detector has essentially the same
  noise level as LIGO 1 for the GW bursts considered here.
   These noise levels are defined so that the
  integrated optimal ( with a matched filter $\propto |f|^{-1/3}$)
  signal to noise ratio (SNR) for each detector is
$SNR =  h^{\rm burst}(f_c)/h^{\rm noise}$. The short-dashed line in the
lower right corner is the   rms  GW amplitude, Eq.(\ref{hrms}).
One sees that the burst amplitude stands 
well above the stochastic background \cite{Nstoch}.
Clearly the search by LIGO/VIRGO of the type of GW bursts discussed here
 is a sensitive probe of the existence of cosmic strings in a larger range of
 values of $\alpha$  than the usually considered search for a stochastic GW background.

\begin{figure}
\vspace{-20mm}
\epsfig{file = 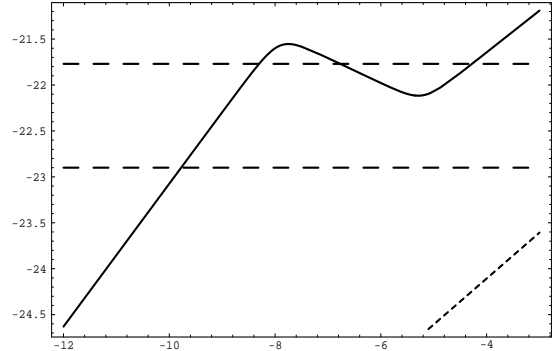,width = 0.4\textwidth, height = 0.6\textwidth,
angle = 0}
\vspace{-20mm}
\caption{Gravitational wave amplitude of bursts emitted by cosmic string
cusps in the LIGO/VIRGO frequency band, as a function of the parameter
$\alpha = 50 G \mu$. [In a base-$10$ log-log plot.]
 The horizontal dashed lines indicate the one sigma noise levels
(after optimal filtering) of LIGO 1 (initial detector) and LIGO 2 (advanced
configuration).The short-dashed line indicates the rms amplitude of the 
stochastic GW background. } 
\end{figure}

  From Fig. 1 we see that the discovery potential of ground-based GW
 interferometric detectors is richer than hitherto envisaged, as it 
could detect cosmic strings in the range $\alpha \agt 10^{-10}$,
i.e. $ G \mu \agt 10^{-12}$. 
Let us also note that the value of $\alpha$ suggested by the
(superconducting-) cosmic-string
Gamma Ray Burst (GRB) model of Ref.\cite{BHV}, namely 
$\alpha \sim 10^{-8}$, nearly corresponds, in Fig. 1, 
to a local maximum of the GW burst amplitude. [This local maximum
corresponds to $z_m \sim 1$. The local minimum on its right corresponds
to $z_m \sim z_{eq}$.] In view of the crudeness of our estimates,
it is quite possible that LIGO 1/VIRGO might be sensitive enough to
detect these GW bursts. Indeed, if one searches for GW bursts which 
are (nearly) coincident with (some\cite{N1}) GRB the needed threshold for a
convincing coincident detection is much closer to unity than in a blind search.
[ In a blind search, by two detectors, one probably needs SNRs $\sim 4.4$
to allow for the many possible arrival times. Note that the optimal filter,
$h^{\rm template}(f) = |f|^{-1/3}$, for our GWBs is parameter-free.]
 
In Fig. 2 we plot $ {\log}_{10}(h^{\rm burst})$ as a function of 
${\log}_{10}(\alpha)$  for 
${\dot N} = 1~{\rm yr}^{-1}$, and for $f = f_c = 3.9 \times 10^{-3}$ Hz.
This frequency is the optimal one for the detection of a  $ f^{-1/3}$ GWB
by the planned spaceborne GW detector LISA. [In determining the optimal
SNR in LISA we combined the latest estimate of the instrumental noise
\cite{schilling} with estimates of the galactic confusion noise \cite{bender}.]
Fig. 2 compares $ h^{\rm burst}(f_c)$  to both LISA's (filtered)
noise level   $h^{\rm noise}$ and to the cosmic-string-generated 
stochastic background $h^{\rm rms}$, Eq.(\ref{hrms}).
 The main differences with the previous
plot are: (i) the signal strength, and the SNR, are typically much higher
for LISA than for LIGO, (ii) though the GW burst signal still stands out 
well above the rms background, the latter is now higher than the (broad-band)
detector noise in a wide range of values of $\alpha$.
LISA is clearly a very sensitive probe of cosmic strings. It might detect
GWBs for values of $\alpha$ as small as
$\sim 10^{-11.6}$. [Again, a search in coincidence
with GRBs would ease detection. Note, however, that, thanks to the lower 
frequency range, even a blind search by the (roughly) two independent arms 
of LISA would need a lower threshold, $\sim 3$, than LIGO.]

\begin{figure}
\vspace{-30mm}
\epsfig{file = 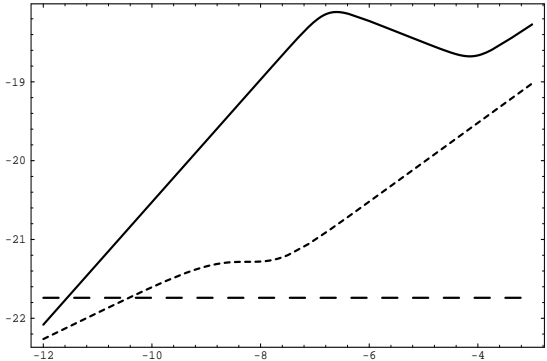,width = 0.4\textwidth, height = 0.6\textwidth, 
angle = 0}
\vspace{-20mm}
\caption{Gravitational wave amplitude of bursts emitted by cosmic string
cusps in the LISA frequency band, as a function of the parameter
$\alpha = 50 G \mu $. [In a base-$10$ log-log plot.] 
The short-dashed curve indicates the rms amplitude of the stochastic GW
 background. The lower long-dashed line indicates the one sigma noise level
(after optimal filtering) of LISA. 
 } 
\end{figure}

We shall discuss elsewhere  the consequences for the interpretation
of the pulsar timing experiments of the
GWB-induced non-Gaussianity of the stochastic GW background \cite{DV}.
 

When comparing our results with observations, one should keep in mind
that the model we used for cosmic strings involves a number of
simplifying assumptions.  {\it (i)} All loops at time $t$ were assumed
to have length $l\sim \alpha t$ with $\alpha\sim \Gamma G\mu$.  It is
possible, however, that the loops have a broad length distribution
$n(l,t)$ and that the parameter $\alpha$ characterizing the typical
loop length is in the range $\Gamma G\mu<\alpha\lesssim 10^{-3}$.
(Here, the upper bound comes from numerical simulations and the lower
bound is due to the gravitational backreaction.)  {\it (ii)} We also
assumed that the loops are characterized by a single length scale $l$,
with no wiggliness on smaller scales.  Short-wavelength wiggles on
scales $\ll \Gamma G\mu t$ are damped by gravitational backreaction,
but some residual wiggliness may survive.  As a result, the amplitude
and the angular distribution of gravitational radiation from cusps may
be modified. {\it (iii)} We assumed the simple, uniform estimate Eq.(\ref{n})
for the space density of loops. This estimate may be accurate in the 
matter era but is probably too small by a factor $\sim 10$ in the radiation
era \cite{Book}. Taking into account this increase of $n_l$ would reinforce
our conclusions in making easier (in some parameter range) the detection
of GWBs.{\it (iv)} Finally, we disregarded the possibility of a nonzero cosmological
constant which would introduce some quantitative changes in our estimates.

The work of A.V. was supported in part by the National Science Foundation.


\begin{thebibliography}{99}
\bibitem{invlength}
 The actual length of the loop changes as the loop
oscillates.  The invariant length is defined as $l=E/\mu$, where $E$
is the loop's energy in its center-of-mass frame.
\bibitem{Nstoch}
The horizontal lines in Fig. 1 (and Fig. 2) should not be used to estimate
the SNR for the detection of the stochastic background, which is optimized
by a different filtering technique.
\end{thebibliography}
\end{document}